\documentclass[10pt,a4paper,notitlepage]{article}

\usepackage{graphicx}% Include figure files
\usepackage{dcolumn}% Align table columns on decimal point
\usepackage{bm}% bold math
\usepackage{subeqnarray}
\date{ }

\begin{document}

\title{\rm \textbf{Properties of gravitational waves in Cosmological General Relativity}  }

\author{\textbf{John G. Hartnett}\\
School of Physics, the University of Western Australia,\\
 35 Stirling Hwy, Crawley 6009 WA Australia\\
\textit{john@physics.uwa.edu.au} \\
\\
\textbf{Michael E. Tobar}\\
School of Physics, the University of Western Australia,\\
 35 Stirling Hwy, Crawley 6009 WA Australia\\
\textit{mike@physics.uwa.edu.au}}

\maketitle

\def\barh{\bar{h}_{\mu \nu}}
\def\barhup{\bar{h}^{\mu \nu}}

\begin{abstract}
The 5D Cosmological General Relativity theory developed by Carmeli reproduces all of the results that have been successfully tested for Einstein's 4D theory. However the Carmeli theory because of its fifth dimension, the velocity of the expanding universe, predicts something different for the propagation of gravity waves on cosmological distance scales. This analysis indicates that gravitational radiation may not propagate as an unattenuated wave where effects of the Hubble expansion are felt. In such cases the energy does not travel over such large length scales but is evanescent and dissipated into the surrounding space as heat.  
\end{abstract}


\section{\label{sec:Intro}Introduction}

In recent decades, the search for gravity waves has intensified with large high powered laser-based interferometic detectors coming on line. See LIGO \cite{LIGO} and TAMA \cite{TAMA} for example. These detectors have already reached sensitivities that should enable them to ``see'' well beyond the local galactic Group. On the other hand, the Hulse-Taylor binary \cite{Taylor1979} ring-down energy budget is a precise test of general relativity and a clear indication of the existence of gravitational radiation, and it seems that the first direct detection is just a matter of time. 

In standard General Relativity the expanding universe has no impact on the properties of gravitational waves, except the the well known effect of redshift. However, in Carmeli cosmology the expansion of the universe (or redshift of the gravitational wave) manifests as a fifth dimension \cite{Carmeli2002c} and in this paper we calculate the effect and how this might impact on a possible direct detection. 
 
\section{\label{sec:CGR}Cosmological General Relativity}

In the late 1990s Moshe Carmeli proposed a new cosmology, Cosmological General Relativity (CGR). \cite{Behar2000, Carmeli1998, Carmeli2002a, Carmeli2002b, Carmeli2002c}  It is a generally covariant theory and extends the number of dimensions of the universe by the addition of a new dimension -- the radial velocity of the galaxies in the Hubble flow. The Hubble law is assumed as a fundamental axiom for the universe and the galaxies are distributed accordingly. 

As a result we have a 5D \textit{spacetimevelocity} universe with two timelike and three spacelike coordinates in the metric. The signature is then $(+\,-\,-\,-\,+)$. The universe is represented by a 5-dimensional Riemannian manifold with a metric $g_{ \mu \nu}$ and a line element $ds^{2}=g_{ \mu \nu}dx^{\mu}dx^{\nu}$. This differs from general relativity in that here the $x^{4} = \tau v$ coordinate is more correctly \textit{velocitylike} instead of \textit{timelike} as is the case of $x^{0} = ct$, where $c$ is the speed of light, a universal constant and $t$ is the time coordinate. In this theory $x^{4} = \tau v$, where $\tau$ is also a universal constant, the Hubble-Carmeli time constant. The other three coordinates $x^{k}, k = 1,2,3$, are spatial and \textit{spacelike}, as in general relativity.

It has been shown that all the results predicted by general relativity and experimentally verified are also predicted by CGR \cite{Carmeli2002b}. However in \cite{Carmeli2002c} Carmeli discussed the one consequence that was not exactly reproduced, and that was gravity waves in 5 dimensions. The new metric resulted in a redshift dependence with a more general wave equation incorporating 5 dimensions $(ct, x^{1}, x^{2}, x^{3}, \tau v)$.

\subsection{\label{sec:linearCGR}Linearized general relativity}

As is the usual practice in a weak gravitational field we write the metric 
\begin{equation} \label{eqn:gmetric}
g_{\mu \nu}=\eta_{\mu \nu} + h_{\mu \nu}
\end{equation}
where the metric $\eta_{\mu \nu}$ is Minkowskian but extended here to 5D from the usual 4D Minkowski metric but with signature $(+\,-\,-\,-\,+)$. Here $\eta_{\mu \nu}$ is perturbed due to gravitating sources with $h_{\mu \nu} \ll 1$. 

A useful tool is to define the trace-reversed $h_{\mu \nu}$ as 
\begin{equation} \label{eqn:hbar}
\bar{h}_{\mu \nu}=h_{\mu \nu}-\frac{1}{2} \eta_{\mu \nu}  h
\end{equation}
where $h = \eta^{\alpha \beta} h_{\alpha \beta}$ is the trace of $h_{\mu \nu}$. Consequently 
\begin{equation} \label{eqn:h}
h_{\mu \nu} = \bar{h}_{\mu \nu}-\frac{1}{2} \eta_{\mu \nu}  \bar{h}
\end{equation}
where $\bar{h} = \eta^{\alpha \beta} \bar{h}_{\alpha \beta}$ and $\bar{h} = -h$.

Then the linearized Einstein field equations to first order in $\bar{h}_{\mu \nu}$ yield
\begin{equation} \label{eqn:linearized}
\bigcirc \bar{h}_{\mu \nu} = -2 \kappa T_{\mu \nu} \qquad  \textrm{plus} \; \eta^{\alpha \beta}\bar{h}_{\mu \alpha, \beta} = 0
\end{equation}
where $\bigcirc$ is the D'Alembertian operator in 5D and may be expressed as
\begin{equation} \label{eqn:DAlamb}
\bigcirc = \left(\frac{1}{c^{2}}\frac{\partial^{2}}{\partial t^{2}} - \nabla^{2} + \frac{1}{\tau^{2}}\frac{\partial^{2}}{\partial v^{2}}\right).
\end{equation}
For conservation of energy and momentum, excluding gravity, it follows from (\ref{eqn:linearized}) that
\begin{equation} \label{eqn:conserved}
\eta^{\alpha \beta}T_{\mu \alpha, \beta} = 0.
\end{equation}

From (\ref{eqn:linearized}) and (\ref{eqn:DAlamb}) it is clear that (\ref{eqn:linearized}) is a generalized wave equation that reduces to 
\begin{equation} \label{eqn:homo}
\bigcirc \bar{h}_{\mu \nu}=0
\end{equation}
in vacuum. 

So the gravitational waves depend not only on space and time but also on the expansion velocity of the source in the Hubble flow. Here $v$ is the fifth co-ordinate, which represents the velocity of the expansion of the space through which the wave passes.

The solution of (\ref{eqn:linearized}) is the sum of the solution of the homogeneous equation (\ref{eqn:homo}), and a particular solution. The following is the special time independent retarded solution in the absence of source-less radiation. The contravariant form is \cite{Ohanian1976}
\begin{equation} \label{eqn:hetro}
\bar{h}^{\mu \nu}= -2\kappa \int \frac{T^{\mu \nu}d^{3}x'}{|\bf{x}-\bf{x'}|},
\end{equation}
where the source mass is located at $\bf{x'}$ and the potential measured at $\bf{x}$. To evaluate the integral in (\ref{eqn:hetro}), provided the measurement point determined by the vector $\bf{x}$ is far away from the source, a Taylor expansion of $1/|\bf{x}-\bf{x'}|$ about $\bf{x'}$ $= 0$ is taken retaining only the first two terms,
\begin{equation} \label{eqn:Taylor}
\frac{1}{|\bf{x}-\bf{x'}|} \approx \frac{1}{r} + \frac{x^{k}x'^{k}}{r^{3}}
\end{equation}
where $r^2=x^{k}x^{k}$.

The integral in (\ref{eqn:hetro}) is then written as
\begin{equation} \label{eqn:hetro2}
\bar{h}^{\mu \nu}= -\frac{GM}{r}-\frac{G}{2}\epsilon^{kln} S^{n}\frac{x^{k}}{r^{3}},
\end{equation}
where the dipole term has been eliminated by choosing the origin of the coordinates to coincide with the center of the source mass. The first term in (\ref{eqn:hetro2}) involves the integral $\int T^{00}(\bf{x'})$$d^{3}x' = M$ identified with the source mass and the second term $\int x'^{k}T^{l0}(\bf{x'})$$d^{3}x' = 1/2\epsilon^{k l n} S^{n}$ where $S^{n}$ is the spin angular momentum of the system. Here $k,l,n = 1, 2, 3$ for the spatial coordinates. 

\subsection{\label{sec:waveeqn}Wave equation in curved \textit{spacevelocity}}

Now considering the time dependence again we can write (\ref{eqn:homo}) as 
\begin{equation} \label{eqn:homo2}
\left(\frac{1}{c^{2}}\frac{\partial^{2}}{\partial t^{2}} - \nabla^{2}\right) \barhup = -\frac{1}{c^2 \tau^{2}}\frac{\partial^{2}}{\partial z^{2}}\barhup,
\end{equation}
where the substitution $v/c \rightarrow z$ has been made. Here $z$ is the redshift of the wave, and the substitution is valid where $z =< 0.1$, which is approximately 400 $Mpc$. We assume it is approximately valid beyond that. Now the solution to (\ref{eqn:homo2}) is the sum of the solution to the homogeneous equation (\ref{eqn:homo2}), which is the usual gravity wave solution in general relativity, and a particular solution of (\ref{eqn:homo2}), which has a redshift dependent source term. 

In fact, in CGR, because the Hubble law is assumed \textit{a priori}, the expansion velocity $v$ (or gravitational wave redshift $z$) is not independent of $r$ and depends on the matter density of the universe. In fact, in the case of CGR, (\ref{eqn:homo2}) can be written as
\begin{equation} \label{eqn:DAlambcurtved}
\left(\frac{1}{c^{2}}\frac{\partial^{2}}{\partial t^{2}} - \nabla^{2} + \frac{1}{c^{2}\tau^{2}}\frac{\partial^{2}}{\partial z^{2}}\right) \barhup = 0,
\end{equation}
with
\begin{equation} \label{eqn:chain}
\frac{\partial^{2}}{\partial z^{2}} = \left\{\left(\frac{\partial r}{\partial z}\right)^{2}\frac{\partial^{2}}{\partial r^{2}}+
\frac{\partial^2 r}{\partial z^2}\frac{\partial}{\partial r}\right\}
\end{equation}
when the chain rule is applied.  

Let us look for a plane wave solution of the form 
\begin{equation} \label{eqn:planewave}
\barhup = \varepsilon^{\mu \nu} \cos k_{\alpha}x^{\alpha},
\end{equation}
where the 3-space co-ordinates are ($x^1$, $x^2$, $x^3$). Here $x^1$ and $x^2$ are orthogonal to the direction of propagation $x^3$ from source to detector. Here $\varepsilon^{\mu \nu}$ is a constant tensor and $k_{\alpha}$ is a constant vector. Therefore we look for a wave propagating in the $r$ direction which would have $k^{\alpha} = (\omega/c,0,0,k_{r})$.

This means we can retain only the $r$ derivative in $\nabla^2$ and effectively re-write (\ref{eqn:DAlambcurtved}) in spherical co-ordinates  as
\begin{equation} \label{eqn:DAlambcurtved2}
\left(\frac{1}{c^{2}}\frac{\partial^{2}}{\partial t^{2}} - \frac{\partial^{2}}{\partial r^{2}} - \frac{2}{r} \frac{ \partial}{\partial r} + \frac{1}{c^{2}\tau^{2}}\frac{\partial^{2}}{\partial z^{2}}\right) \barhup  = 0.
\end{equation}

Eqs (\ref{eqn:DAlambcurtved}) and (\ref{eqn:DAlambcurtved2}) are only valid where the Hubble law applies. When it doesn't apply, that is, where $\partial r/\partial z =0$ in (\ref{eqn:chain}), (\ref{eqn:DAlambcurtved}) becomes the normal wave equation for gravity waves in source free regions. 

However where the Hubble law is applicable, in flat (i.e. $\Omega = 1$) \textit{spacevelocity} $\partial r/\partial z = c \tau$, which is the Hubble law in the zero distance/zero gravity limit. In the general curved \textit{spacevelocity}, the form of the derivative is given by \cite{Hartnett2005}
\begin{equation} \label{eqn:deriv}
\frac{1}{c^{2} \tau^{2}}\left(\frac{\partial r}{\partial z}\right)^{2} = 1+ (1-\Omega) \frac{r^{2}}{c^{2} \tau^{2}}, 
\end{equation}
where $\Omega = \rho/\rho_{c}$ is the mass/energy density at some epoch expressed as a fraction of the `critical' density, $\rho_{c} = 3/8\pi G \tau^{2}$.

Substituting (\ref{eqn:deriv}) into (\ref{eqn:DAlambcurtved2}) with (\ref{eqn:chain}) we get
\begin{equation} \label{eqn:DAlambcurtved3}
\left(\frac{1}{c^{2}} \frac{\partial^{2}}{\partial t^{2}}- \frac{2}{r} \frac{ \partial}{\partial r} + \frac{1-\Omega}{c^{2}\tau^{2}}\left\{r^{2}\frac{\partial^{2}}{\partial r^{2}}+r\frac{\partial}{\partial r}\right\} \right) \barhup = 0,
\end{equation}
which only has dependence on $r$ and $t$. This is a new equation depends on the surrounding matter density $\Omega$. 

When the gravity wave is very distance from the source and when $r \gg c\tau/\sqrt{1-\Omega}$ the second term of (\ref{eqn:DAlambcurtved3}) is much smaller than the term in curly brackets, we assume the second term negligible.  Therefore (\ref{eqn:DAlambcurtved3}) can be approximated for large $r$ as
\begin{equation} \label{eqn:DAlambcurtved4}
\left(\frac{1}{c^{2}} \frac{\partial^{2}}{\partial t^{2}} + \frac{1-\Omega}{c^{2}\tau^{2}}\left\{r^{2}\frac{\partial^{2}}{\partial r^{2}}+r\frac{\partial}{\partial r}\right\} \right) \barhup = 0.
\end{equation}

The solution of (\ref{eqn:DAlambcurtved4}) can be obtained by separation of variables assuming a solution of the form $\barhup \propto R(r)e^{i (\omega t+k_z)}$. Substituting the latter into (\ref{eqn:DAlambcurtved4}) yields   
\begin{equation} \label{eqn:solnwavenumber}
\frac{\omega^{2}}{c^{2}} +\frac{\Omega - 1}{c^{2}\tau^{2}} = 0,
\end{equation}
with $k_z \approx 0$ and $R(r) = a_{1}r^{-1}$ for $r \gg c\tau/\sqrt{1-\Omega}$. More generally $R(r) = \sum a_{n}r^{-n}$ a polynomial expression with an index $n > 0$.  Furthermore by taking a hint from the solution of the usual heterogeneous equation shown in (\ref{eqn:hetro2}) $R(r)$ is determined as
\begin{equation} \label{eqn:solnR}
R(r)= -\frac{GM}{r}- \mathcal{O}(\frac{1}{r})^{3}.
\end{equation}
Equation  (\ref{eqn:solnwavenumber}) is a resonance condition. For this solution, which spans the whole extent of the Universe, the Universe acts like a resonant mode with a characteristic scale radius \cite{Oliveira2005a, Oliveira2005b} of 
\begin{equation} \label{eqn:scaleR}
R_{\Omega}= \sqrt{|R_{\Omega}^2|}=\sqrt{|\frac{c^2}{\omega^2}|}=\frac{c\tau}{\sqrt{|1-\Omega|}},
\end{equation}
and resonance frequency 
\begin{equation} \label{eqn:freq}
\omega = \frac{\sqrt{1-\Omega}}{\tau}.
\end{equation}
For values of $\tau = 4.2 \times 10^{17}\;s$ and $\Omega = 0.02$ the scale radius is $R_{\Omega} \approx 4.13 \,Gpc$ and the characteristic frequency is $\omega/2\pi \approx 3.66 \times 10^{-19}$ Hz. 

When $r \approx < c\tau/\sqrt{1-\Omega}$ the second and fourth terms of (\ref{eqn:DAlambcurtved3}) are much smaller than the third, and hence can be neglected.  Therefore (\ref{eqn:DAlambcurtved3}) can be approximated as
\begin{equation} \label{eqn:DAlambcurtved5}
\left(\frac{1}{c^{2}} \frac{\partial^{2}}{\partial t^{2}} + \frac{1-\Omega}{c^{2}\tau^{2}}r^{2}\frac{\partial^{2}}{\partial r^{2}} \right) \barhup = 0.
\end{equation}

Now for a spherically symmetric expanding universe Carmeli \cite{Carmeli2002a,Hartnett2005} obtains the relation between redshift and distance of the emitting source, 
\begin{equation} \label{eqn:rz}
\frac{r}{c \tau} = \frac{\sinh(\varsigma \sqrt{1-\Omega})}{\sqrt{1-\Omega}}, 
\end{equation}
where $\varsigma = ((1+z)^2-1)/((1+z)^2+1)$. Assuming this relation also holds for gravity waves  (\ref{eqn:DAlambcurtved4}) becomes
\begin{equation} \label{eqn:DAlambcurtved6}
\left(\frac{1}{c^{2}} \frac{\partial^{2}}{\partial t^{2}} + \sinh^2(\varsigma \sqrt{1-\Omega})\frac{\partial^{2}}{\partial r^{2}} \right) \barhup = 0, 
\end{equation}

\begin{figure}
\includegraphics[width = 3.5 in]{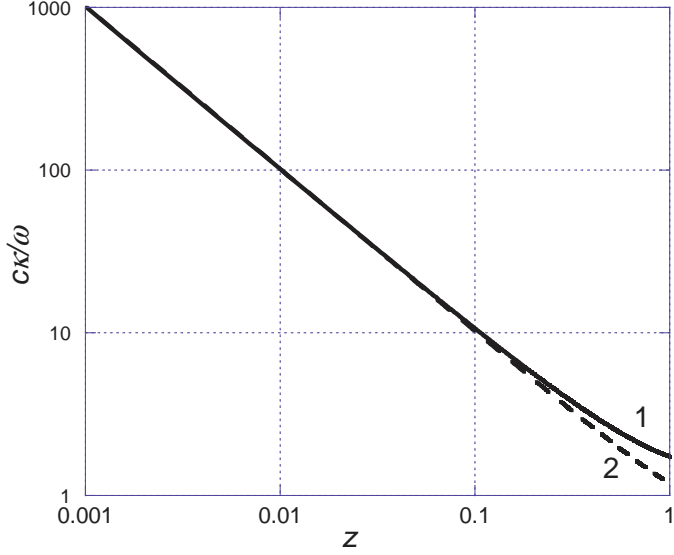}
\caption{\label{fig:fig1} The decay constant $c \kappa /\omega$ as function of redshift, $z$ shown from Eq. (\ref{eqn:dispersion}) as the solid curve 1 and from the approximated Eq. (\ref{eqn:dispersion2}) as the broken curve 2. $|c \kappa /\omega|$ asymptotes to unity in the limit as $v \rightarrow c$ or when $z \rightarrow \infty$. The values of $\Omega$ changes from $\Omega <1$ to $\Omega > 1$ at some value of $z$ but the sinh function is continuous changing into a sine function as the argument of the sinh function changes from real to imaginary}
\end{figure}

This is now a wave equation with a solution of the form $\barhup \propto e^{i (k_{r}r+\omega t)}$, which results in the following dispersion relation
\begin{equation} \label{eqn:dispersion}
k_{r}^{2} \approx -\frac{\omega^{2}/c^{2}}{\sinh^2(\varsigma \sqrt{1-\Omega})},
\end{equation}
which can be approximated for $z \ll 1$ as
\begin{equation} \label{eqn:dispersion2}
k_{r}^{2} \approx \frac{\omega^{2}/c^{2}}{(\Omega-1)z^{2}}.
\end{equation}
When $\Omega > 1$ the wave number is real and approximately $k_{r}=\omega/(c z \sqrt{\Omega - 1})$ and hence gravity waves propagate yet are dependent on redshift. When $\Omega < 1$ the wave number is imaginary and the amplitude is attenuated with a decay constant $\kappa = ik_{r}= \omega/(c z \sqrt{1-\Omega})$. In Fig. \ref{fig:fig1} we plot the decay constant $\kappa$ normalized by $\omega/c$, using $k_{r}$ from (\ref{eqn:dispersion}). From the figure it is apparent that the approximation of (\ref{eqn:dispersion2}) is good for $z < 0.1$.

It follows from (\ref{eqn:dispersion}) the phase and group velocities are determined from
\begin{equation} \label{eqn:velocity}
\frac{\partial \omega}{\partial k_{r}} = \frac{\omega}{k_{r}}= c\sqrt{-\sinh^2(\varsigma \sqrt{1-\Omega})}.
\end{equation}
Using the identity $i\sin \theta = \sinh (i\theta)$ (\ref{eqn:velocity}) becomes
\begin{equation} \label{eqn:velocity2}
\frac{\partial \omega}{\partial k_{r}} = \frac{\omega}{k_{r}}= c \sin(\varsigma \sqrt{\Omega-1}),
\end{equation}
where $\Omega \geq 1$. The latter can be approximated for $z \ll 1$ as $c z \sqrt{\Omega-1}$. In the general cosmos where $\Omega > 1$ gravity wave propagate with the velocity $c \sin(\varsigma \sqrt{\Omega-1}) \rightarrow c$ where $v \rightarrow c$. In the cosmos where $\Omega <1$ we have evanescent decay.

The dependence in (\ref{eqn:deriv}) is not the same within a bound galaxy of stars and gas as it is for the large scale structure of the expanding universe, which considers only the center of mass motion of galaxies within it. This is because within a galaxy (or cluster) the full effect of the Hubble expansion is not felt with respect to the center of mass, of which the gravitational radiation must travel with respect to, and here we show it results in the same solution as standard General Relativity. 

For spherically symmetric distribution of matter in a galaxy with the region of interest far from the central potential of a fixed mass, in the disk region, it may be derived (Eqs B.63a and B.67 of Carmeli \cite{Carmeli2002a}) that 
\begin{equation} \label{eqn:derivGalaxy}
\frac{1}{c^{2} \tau^{2}}\left(\frac{\partial r}{\partial z}\right)^{2} = (\Omega-1) \frac{r^{2}}{c^{2} \tau^{2}}, 
\end{equation}
where $z = v/c$ and $\Omega$ is the mass density, with the origin of coordinates coinciding with the origin of the spherically symmetric gravitational potential.

Using (\ref{eqn:derivGalaxy}) in (\ref{eqn:DAlambcurtved2}) with (\ref{eqn:chain}) results in a wave equation
\begin{equation} \label{eqn:DAlambcurtved7}
\left(\frac{1}{c^{2}} \frac{\partial^{2}}{\partial t^{2}}- \frac{ \partial^2}{\partial r^2}-\frac{2}{r} \frac{ \partial}{\partial r} + \frac{\Omega-1}{c^{2}\tau^{2}}\left\{r^{2}\frac{\partial^{2}}{\partial r^{2}}+r\frac{\partial}{\partial r}\right\} \right) \barhup = 0,
\end{equation}
where we can neglect the terms in the curly brackets because they are insignificant on the scale of a galaxy. Also after neglecting the third term for distant sources, we get the normal gravity wave equation of GR, that is,
\begin{equation} \label{eqn:DAlambcurtved8}
\left(\frac{1}{c^{2}} \frac{\partial^{2}}{\partial t^{2}}- \frac{ \partial^2}{\partial r^2} \right) \barhup = 0.
\end{equation}

\subsection{\label{sec:scales}Density scales in the universe}

On the local scale the Hubble law $v = r/ \tau$ does not apply or is so insignificant as to be negligible. On that scale therefore effects of \textit{spacevelocity} are negligible, or in other words, $dv \rightarrow 0$. That is the realm where GGR reduces to the usual special and general relativity theory. Gravity waves propagate as is usually expected according to (\ref{eqn:DAlambcurtved8}). 

On the cosmological scale we expect the Hubble law to be very significant and hence that is the realm where \textit{spacevelocity} is operative. On that scale we can now also consider (\ref{eqn:DAlambcurtved5}) to represent a genuine modification to the usual 4D \textit{spacetime} equation found in general relativity textbooks. The effect of \textit{spacevelocity} is contained in the modified D'Alembertian operator. This means close to the source gravitational energy is emitted in the usual fashion as described by (\ref{eqn:hetro2}). Over cosmological length scales, however, CGR predicts the gravitational waves from distant galaxies will be fully attenuated by the time the reach the Earth. Thus, projects such as the Large-scale Cryogenic Gravitational-wave Telescope (LCGT) in Japan could be very important to test CGR theory \cite{Kuroda}. For example this project will detect gravity-waves from coalescing neutron-star binary systems 200 $Mpc$ away at a SNR of 10. In contrast the TAMA field of view is 1 $Mpc$ and LIGO is 20 $Mpc$. Thus, if the LGCT field of view includes a much larger volume of the Universe than LIGO or TAMA, and a smaller event rate than predicted by GR, this could be evidence for CGR.

The best estimates of the local baryonic matter density puts it around $\Omega_{m} \approx 0.02$ \cite{Fukugita1998} for the present epoch. As has been shown \cite{Hartnett2004a, Hartnett2005}, using the Carmeli theory it is not necessary to assume any dark matter in the cosmos, therefore the matter density between galaxies should be $\Omega < 1$ even out to a redshift $z = 2$ \cite{Hartnett2005}. Equation (\ref{eqn:deriv}) is valid for large $z$ therefore the analysis applies.

Accordingly the averaged matter density of the universe has been determined at the current epoch $\Omega_m = 0.021 \pm 0.042$ \cite{Hartnett2005} and can be approximately related by $\Omega=\Omega_m (1+z)^3$ as a function of redshift for $z \leq 1$. Using this relation and the form of (\ref{eqn:dispersion}) the decay constant ($\kappa$) is shown as $c \kappa /\omega$ in Fig. \ref{fig:fig1} as a function of redshift, $z$. The value of $c \kappa /\omega$ in Fig. \ref{fig:fig1} is limited by the unknown form of $\Omega(z)$ for $z \gg 1$. However, over the epochs shown, CGR predicts that gravitational waves do not propagate at scales beyond galaxies (and clusters).

Gravity waves that are generated within a galaxy quickly decay in the void between. Gravitational radiation therefore leaks into the surrounding space according to (\ref{eqn:DAlambcurtved5}) with attenuated amplitudes when $\Omega$ drops below unity. According to CGR, gravity waves will not propagate far in an expanding universe.	Therefore we would expect to see no stochastic gravity wave background spectrum. Instead we conjecture that the energy is deposited into space as heat. As a result they may contribute to the CMB blackbody temperature.	 

The binary pulsar PSR B1913+16, discovered in 1974 by Russell Hulse and Joseph Taylor \cite{Hulse1975}, for which they won the 1993 Nobel prize, consists of two neutron stars closely orbiting their common center of mass. One of them is a pulsar with a rotational period of 59 ms and extremely stable compared to other pulsars. The two neutron stars slowly spiral toward their common center of mass radiating energy. The orbit period however is declining by about $7.5 \times 10^{-5}$ seconds per year on an orbit period of $7.75$ hours \cite{psr1913}. This change, believed to result from the system emitting energy in the form of gravitational waves, has been a very precise  and successful test of general relativity \cite{Taylor1979}. 
 
On the scale of the Galaxy it expected that there are still some small effects of the Hubble law \cite{Hartnett2004b, Hartnett2005b}. These effects modify the dynamics of the motion of tracer gases in the outlying regions of the Galaxy but in regard to gravity waves in galaxies they propagate unhindered, as in normal GR. Therefore it is expected that the energy dissipated by the Hulse-Taylor binary does travel as gravity waves within the Galaxy, but will be attenuated outside the region of the Galaxy where the mass density $\Omega$ drops below unity.

\section{\label{sec:Conclusion}Conclusion}

This paper derives the wave propagation equation for gravitational radiation in an expanding universe where an additional constraint has been placed on the nature of space itself. This is the introduction into the metric the fundamental assumption of the expansion of space according to the Hubble law. It is then found that no unattenuated gravity wave  propagation may be possible in regimes where the Hubble expansion has effect. Then, depending on the density of matter, the propagation constant of any gravitational wave is either real or imaginary. If imaginary it represents an evanescent wave, which we conjecture means the energy is dissipated into the surrounding space as heat. When gravity waves are eventually detected, a test of this theory would be the detection of gravity waves from within the Galaxy but not from extra-galactic sources.

\section{Acknowledgment}
The authors wish to thank Dr Paul Abbott and Prof. Bahram Mashhoon for their helpful advice and suggestions.

\end{document}